\pdfoutput=1

\newif\ifpublic
\publictrue

\documentclass[a4paper]{llncs}

\usepackage[utf8]{inputenc}

\usepackage[giveninits=true,url=false,doi=false,isbn=false,backend=biber,maxbibnames=9,maxcitenames=9,]{biblatex}
\usepackage{url}
\usepackage{placeins}

\usepackage[pdftex]{graphicx}
\usepackage[a4paper, margin=2.8cm]{geometry}

\usepackage{listings}
\usepackage{float}
\floatstyle{ruled}
\newfloat{listing}{h}{lop}
\floatname{listing}{Listing}

\usepackage[colorlinks, linkcolor=blue, urlcolor=blue, citecolor=blue]{hyperref}

\begin{document}

\title{Side-channel based intrusion detection for industrial control systems}

\ifpublic
\author{
Pol Van Aubel\inst{1}
\and
Kostas Papagiannopoulos\inst{1}
\and
\L{}ukasz Chmielewski\inst{2}
\and
Christian Doerr\inst{3}
}

\institute{
Radboud University, Digital Security Group,
\email{\{pol.vanaubel,k.papagiannopoulos\}@cs.ru.nl}
\and
Riscure BV, Delft, the Netherlands,
\email{Chmielewski@riscure.com}
\and
Delft University of Technology, Department of Intelligent Systems,
\email{c.doerr@tudelft.nl}
}
\fi


\maketitle

\begin{abstract}
Industrial Control Systems are under increased scrutiny. Their security is
historically sub-par, and although measures are being taken by the
manufacturers to remedy this, the large installed base of legacy systems
cannot easily be updated with state-of-the-art security measures.
We propose a system that uses electromagnetic side-channel measurements
to detect behavioural changes of the software running on industrial control systems.
To demonstrate the feasibility of this method, we show it is possible to
profile and distinguish between even small changes in programs on
Siemens S7-317 PLCs, using methods from cryptographic
side-channel analysis.

\end{abstract}

\ifpublic
\catcode`@=11 \def\@thefnmark{} \@footnotetext{This work was supported
by the Dutch electricity transmission system operator TenneT TSO B.V.
\\
Permanent ID of this document: {\tt b7751e7466fc1c298da31a840c902cbc}.
Date: 2017-12-14}
\fi

\setlength{\textfloatsep}{8pt plus 1.0pt minus 2.0pt}

\section{Introduction}
Industrial control systems (ICS) are used to manage most of our
critical infrastructures. With the move toward more centralized
control using IP-based networks, these systems, which historically
have not needed advanced protection mechanisms, are opened to a
wider range of attack scenarios. One such scenario is an attacker
modifying the software running on the system, e.g. to perform a
long-running attack on the industrial process being controlled,
as happened with Stuxnet in the uranium enrichment facilities in
Natanz; or in preparation for a later, sudden attack that takes
down a significant part of the electricity grid, as happened in
Ukraine in 2015 and 2016.

In general, an operator of ICSs would like to prevent compromised
software from being installed. Solutions for this can be found in software
integrity verification, and software inspection.
Software integrity can be determined by e.g. taking a signed software
image and verifying the signature with a trusted platform module.

Prevention of system compromise through software inspection is a
technique widely used, with varying success, in the IT landscape.
There exists a variety of intrusion detection \& prevention systems that are capable of monitoring
the network or the host systems themselves~\cite{snort,adaptbroscada,ossec}.
To actively prevent ICS compromise during an attack,
these systems can e.g. stop communication between the attacker
and the ICS, or stop execution of the software under attack. However, this
requires software integration in the monitored ICS, which is not always
a feasible option for existing legacy systems, and comes with other drawbacks
such as influencing the characteristics of the system being monitored.

Even if these solutions are available and effective in preventing compromised
software from running, uncompromised software may
still be made to misbehave. Bugs in the software or compromise of the underlying
system can allow an attacker to circumvent prevention mechanisms.
Detecting this situation is an important part of
any system intended to defend ICSs against attackers. In this paper, we
will focus on this \emph{detection}, rather than prevention. Specifically, we
attempt to detect changes in the behaviour of software.

Detecting the anomalous behaviour
that compromised software exhibits becomes harder when the system running
that software behaves unpredictably to begin with. This is
often the case for systems with a lot of human interaction.
However, ICSs are inherently more stable and predictable, making our task
easier. Nonetheless, detecting software compromise on ICSs is not
straightforward.

Proposed methods for detecting software compromise often rely on non-existent
hardware support, instruction set modifications, operating system
modifications, etc.~\cite{SCHEM,zhang2004hardware,anomalouspathdetection}
These are all unavailable for the huge number
and wide range of control systems currently deployed in the world's
heavy industry and critical infrastructure. Symbiotes, proposed by
Cui and Stolfo\cite{Cui2011},
try to remedy this by offering a general solution that allows retrofitting
defensive software in existing firmware images. The exact functionality
of the original firmware does not need to be known for this, which means
the technique can be applied to a wide range
of embedded systems. However, it does require changing the original
manufacturer-provided firmware image, and might therefore not be an
acceptable solution for many operators of ICSs.

Detection systems running on the ICS itself
may not be able to detect all targeted attacks. For instance,
Abbasi and Hashemi have shown that it is possible to circumvent
existing host-based intrusion detection with an attack that 
reconfigures a Programmable Logic Controller's (PLC) processor pin configuration on-the-fly~\cite{eemcs27470}.
Another attack that may not be detected is the complete
replacement of a device's firmware~\cite{cui2013firmware,BASNIGHT201376,peck2009leveraging},
since the detection system is part of that. Indeed, the threat model
of Symbiotes explicitly excludes the replacement of the entire firmware
image~\cite{Cui2011}.

\vspace{-.5em}
\subsubsection{Our Contribution.} In this work, we propose an alternative approach to detecting software compromise
which uses side-channel measurements of the underlying hardware. Side-channel analysis
is a common technique in security evaluations, since it can be used to
distinguish system behaviour that differs slightly based on some secret information such as
a cryptographic key.
We posit that similarly, it is possible to use side-channels to verify that software is still
behaving as intended, based on some baseline of behaviour. 
Our approach using side-channels has the advantage that there is no need for
monitoring support in the device firmware, and, by extension, that it will continue to
function if the device is compromised.
Our contributions are as follows:
\begin{enumerate}
	\item We verify the applicability of a side-channel-based intrusion detection system (IDS) in a
		real-world scenario, using measurements of the electromagnetic
		(EM) emissions from the processor on a Siemens Simatic S7-317
		Programmable Logic Controller (PLC).
		
	\item We describe in detail how to deploy such an IDS,
		highlighting its modus operandi, the adversarial model
		considered and the necessary modifications to the existing ICS
		hardware.

	\item We suggest a two-layer intrusion detection strategy that can
		effectively detect the illegitimate behaviour of a user program
		(part of the software running on a PLC),
		even when only minor malicious alterations have been performed. We
		describe the statistical models that profile the user program
		and demonstrate how side-channel emission templating is
		directly applicable in the IDS context.
\end{enumerate}

\vspace{-.5em}
\subsubsection{Related Work.}
Side-channel-based techniques are becoming an increasingly popular tool to
verify software, as suggested by Msgna et al.~\cite{Msgna2014-verifying} and
Yoon et al.~\cite{Yoon:2015:MHM:2744769.2744869}. Similarly, Liu et
al.~\cite{DBLP:conf/ccs/LiuWZZXX16} managed to perform code execution tracking
and detect malicious injections via the power side-channel and a hidden Markov
model. In a hardware-oriented scenario, Dupuis et
al.~\cite{DBLP:journals/isjgp/DupuisNFR13} have used side-channel approaches in
order to detect malicious alterations of integrated circuits, such as hardware
trojan horses. In the field of reverse engineering, work by
Goldack~\cite{goldack2008side}, Eisenbarth et al.~\cite{Eisenbarth2010},
Quisquater et al.~\cite{Quisquater:2002:ACR:1250988.1250994} and Vermoen et
al.~\cite{Vermoen2007} has shown the feasibility of using power traces to
reverse-engineer software, reaching instruction-level granularity. More
recently, Strobel et al. have shown that EM emissions can similarly be used for
reverse engineering purposes~\cite{7092372}.

Previous works attempt detection at various levels of granularity ranging from
recognizing single instructions to detecting larger blocks. In our
work, we demonstrate that using EM emissions as a mechanism to detect software
compromise is possible without mapping the observed measurements to specific
instructions, or indeed even knowing the instruction set of the chip being
monitored. Our analysis is carried out on a processor that is part of a larger
PLC, deployed in many systems around the world. In particular, we do not
control the clock speed, cannot program the processor directly with its
low-level instruction set, and cannot predict its behaviour with
regards to EM emissions beforehand.

\section{Software Behaviour Verification on Programmable Logic Controllers}
In section~\ref{sec:PLCs}, we briefly describe the general architecture of PLCs, and
explain why they are particularly suited for the approach
we propose. Next, we introduce the EM side-channel in section~\ref{sec:emsca}.
Then, in sections~\ref{sec:model} and~\ref{sec:twolayer}, we
describe our attacker model and propose a two-layer IDS strategy
that employs the EM leakage to perform behavioural verification. We also
describe the required PLC modifications to apply the IDS to legacy systems.
Finally, in section~\ref{sec:operation} we highlight the operation of
our system.

\subsection{Programmable Logic Controllers}
\label{sec:PLCs}
A Programmable Logic Controller (PLC) is an industrial computer designed for highly
reliable real-time measurement and control of industrial processes. PLCs are designed
to be easy to program, and in their most basic function simply emulate a logic network
that reads inputs and drives outputs based on the values of those inputs. The operator
of a PLC creates a program, which we will call \emph{``user program''}, to perform this
control. A modern PLC runs some version of a real-time operating system (OS), which provides
functionality such as network connectivity to other machines, communication bus control,
reading inputs into memory, driving outputs from memory, and running the user program.
The latter three form the Read-Execute-Write (REW) cycle.

During the run of the user program, most low-priority tasks such as network
communication are postponed. This is to guarantee a maximum execution time on the program,
offering real-time guarantees to the operator. This means that in theory, the
execution of a user program is not often preempted by other code, and it should therefore
be relatively easy to observe the behaviour of the user program and determine whether it
is, in fact, still behaving the way it should be.

Doing this observation from within the PLC itself is not trivial and requires extensive
modifications to their OS. Even though support of the PLC vendor is not always
required for this \cite{Cui2011}, it is unclear whether it would be wise to
modify the OS on existing PLCs, because it introduces concerns such as the
possibility of breaking real-time guarantees.

\begin{figure}[t]
	\centering
	\includegraphics[width=.9\textwidth,height=.18\textheight]{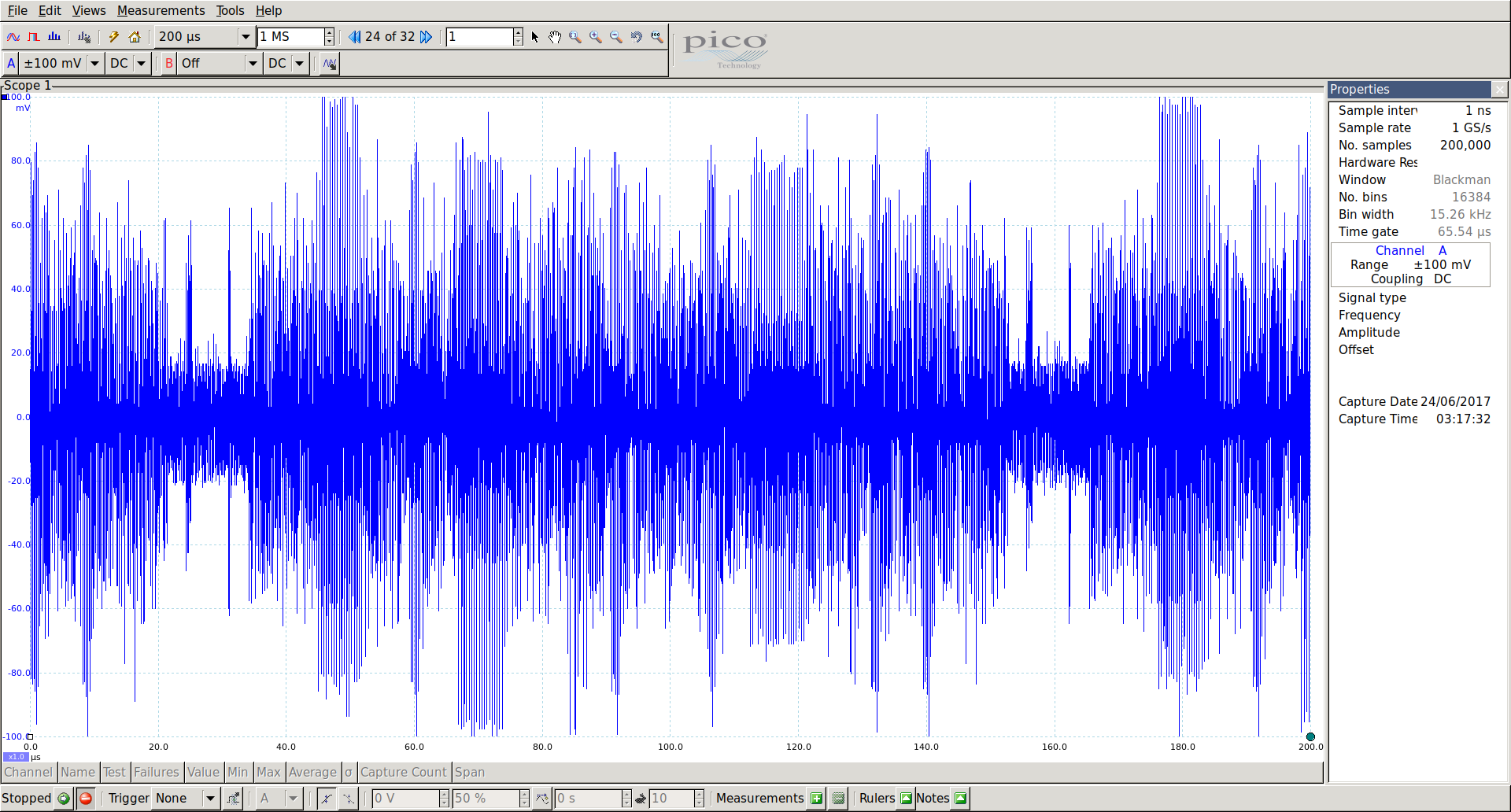}
	\caption{EM radiation captured from a running Siemens S7-317 PLC}
	\label{fig:emtrace}
\end{figure}

\subsection{EM Side-Channel Analysis}
\label{sec:emsca}
To enable us to still observe the user program in a less intrusive manner, we
consider a concept used in cryptanalysis to observe and break cryptographic
implementations, namely side-channel leakage. A side-channel can be thought of as a non-functional
transmission of information about the state of a system. E.g., the temperature of a processor
is not a functional aspect of it, but its level of activity can easily be derived from it\footnotemark.
Silicon chips emit electromagnetic (EM) radiation caused by the electrical characteristics
of the operations they perform.
This radiation can be captured using an EM-probe,
basically a looped wire responding to changes in the EM field it resides in, connected
to a high-speed oscilloscope. 
Figure~\ref{fig:emtrace} shows a capture of EM radiation from the control chip
of a PLC, revealing when the OS, user program, and specific blocks of operations in the
user program are executed, and showing clear regularity.

\footnotetext{TEMPEST is an NSA program dealing with spying on information systems through the
use of these side-channels.}

Side-channel leakages are most commonly used in the analysis of cryptographic
hardware such as smart cards; a large body of research exists that shows how to
extract cryptographic keys from otherwise protected devices, using
sophisticated EM
techniques~\cite{DBLP:journals/integration/PeetersSQ07,DBLP:journals/iacr/GaleaMPT15,DBLP:conf/ctrsa/HeyszlMHSS12}.
More interestingly, the side-channel literature has established a wide spectrum
of \emph{templating} techniques, i.e. statistical models that, once
sufficiently trained, can help us distinguish between different states of  a
system~\cite{DBLP:conf/ches/ChariRR02,DBLP:conf/cardis/ChoudaryK13,DBLP:conf/ches/SchindlerLP05}.
Our work employs such templating techniques to provide intrusion detection
capabilities.

\subsection{Attacker model}
\label{sec:model}
Our system is intended to defend against an attacker who can upload new software
to the PLC to replace
or modify the existing user program. The attacker does not control the PLC
operating system. Although this is not a very strong attacker model, it is
a realistic one. Public analysis of Stuxnet has revealed that it functioned
by replacing the user program on the PLCs it targeted~\cite{symantic-stuxnet},
which means it falls within our attacker model.
However, the more recently revealed Industroyer malware~\cite{industroyer,crashoverride}
does not modify software on a PLC, and therefore does not fall within our attacker model.

\subsection{User Program Intrusion Detection System}
\label{sec:twolayer}
We propose a two-layer intrusion detection system (IDS) 
that uses this EM side-channel to verify that a PLC's user program
is still behaving the way it was programmed to behave.
For this, it is not necessary to know
which exact operations a chip is performing; only that they are still the same
based on some baseline profile established
in the past. The IDS would record this profile when the PLC is first deployed,
and it should be updated whenever legitimate code changes are performed.

To verify that the user program behaves as expected,
the system uses the following two layers of verification, alerting the operator
as soon as one layer shows compromise.
\begin{enumerate}
	\item The first layer checks user program runtime. If the user program
		deviates in runtime, this is a clear indicator that it is not
		behaving as intended. If the runtime is
		deemed to be close enough to potentially be legitimate, the
		system checks the second layer.

	\item For the second layer, the user program's EM trace is compared to
		a baseline profile that has been crafted by templating the
		emitted side-channel leakage. If it matches sufficiently, the
		software is behaving legitimately.
\end{enumerate}

Extensive malicious alterations by an adversary unaware of this system are
easy to detect via layer 1. If an adversary is aware of the functioning of the
system, or by coincidence happens to craft a user program that runs in the
same amount of time, they will be detected by layer 2.

\subsubsection{Layer 1: Timing Side-Channel.}
Program runtime can be determined either by the PLC informing the IDS when it hands
over execution to the user program, and when it regains control (we call this a
trigger signal); or by analysing the EM waveform to spot when the control handover
happens.

\label{sec:triggers}
\begin{enumerate}
	\item Trigger signals can be used by the monitoring oscilloscope to
		know when to capture the EM waveform. The PLC's operating
		system could send such a signal every time it hands over
		execution to the user program, and drive the signal low again
		once it regains control.  The advantage of this is that the
		oscilloscope always captures the exact waveform that we are
		interested in, without any need for the post-processing
		described in
		section~\ref{sec:construction}. This does require a modification to the PLC
		operating system, however, which may not always be possible.
		Obviously, the emission of this signal should not be blockable
		from the user program logic, and so could also require the
		addition of a hardware output that cannot be driven from the
		user program.

	\item Waveform analysis uses the same EM side-channel as layer 2, described below, for matching
		the user program: an oscilloscope can simply capture long runs
		of the complete EM waveform, both OS and user program
		emissions. These waveforms can then be searched for some
		profiled parts of the operating system known to be right before
		the start and right after the end of the user program.
\end{enumerate}

\subsubsection{Layer 2: EM Side-Channel.}
It is not straightforward to distinguish user program compromise from other deviations from the norm:
there is the case where the controlled industrial
process goes outside of its target values, and needs to be corrected; or the case where
a very infrequent but legitimate action is taken, such as opening a breaker in a power
distribution grid.
At that point, the user program's behaviour will deviate from
the norm,
but
we should not alert when it happens.
This shows that it
is not sufficient to profile only the common case; the user program must
be profiled under each combination of inputs that leads to a different
path through the program.

When the user program actually behaves differently than intended, either
through misconfiguration, bugs, or malicious intervention, these
unintended deviations from the norm should all be detected. This means our
problem is to reliably distinguish between:
\begin{itemize}
	\item when the user program is running in one of its usual paths;
	\item when the same user program is taking a legitimate, yet unusual
		path;
	\item when something other than a legitimate path is taken, or
		another user program is running.
\end{itemize}
Distinguishing between the first two cases is not strictly necessary for our IDS, but may
be useful for checking if the legitimate but unusual path is taken under the correct
conditions. We split this problem into four distinguishing cases:
\begin{enumerate}
	\item Can we reliably distinguish user program A from user program B?
	\item Can we reliably distinguish between different paths in the same
		user program?
	\item Can we reliably distinguish paths in user program A from paths in user program B?
	\item Can we reliably recognize whether a user program is user program A or not?
\end{enumerate}

Question 4 is not
strictly a distinguishing case. Instead, we need a threshold beyond which we no longer
accept a program as being program A. We determine this threshold
experimentally using a few programs with minor modifications.
Different user programs might require a different threshold,
and determining such a threshold would be part of any profile building effort.

\subsubsection{PLC modifications.}
Our system does require one modification to existing hardware:
the processor needs to be fitted with an \emph{EM sensor}. Although this
might imply that our technique is, in fact, not applicable to existing legacy
systems, we believe that this is a modification that can realistically be performed on existing
systems, without support from the vendor. The exact
location of the sensor depends on the location of the processor executing the
user program; in general, the sensor would be a loop situated right
on top of the processor. Stable orientation of the sensor would require it to either
be fixed in place using e.g. hot glue, or use of a bracket mounted on the
external housing of the PLC, with the sensor inserted through ventilation
grating.

\subsection{Operation}
\label{sec:operation}
There are two ways to use our IDS: first, constant operation,
where the EM side-channel is constantly monitored and checked for anomalies; and second,
spot-checks, where an engineer manually attaches monitoring equipment every
so often which then checks whether the PLC is behaving satisfactory. 
Considering the potential cost of the monitoring equipment, in particular the
high-speed oscilloscopes, spot-checks seem the more likely way to use our system.

A smart attacker could try to hide in the periods between spot-checks.
However, consider that for an attacker to hide their presence, they would need
logic to determine whether a check is happening. The execution of this logic is
detectable by our IDS.  The same is true for a dormant backdoor, since it must
contain logic to check whether it should start executing.

%


\section{Experimental Setup}
We experimentally verify the feasibility of our proposed system.

Our main experiment setup consists of a Siemens S7-317 PLC, modified
to enable it to run outside of its casing. We use a PCBGRIP~\cite{pcbgrip}
kit to hold both the main PLC board and the probe, so that any disturbances
do not move the probe relative to the chip under test.

\subsection{Measurement setup}
We measure EM radiation of the PLC's main processor, an
Infineon Tricore SAFTC11IA64D96E\footnotemark, using a
Langer RF-R 50-1 10mm loop probe, which has a frequency range
of 30MHz -- 3GHz. The probe is connected to a DC-powered Riscure amplifier
with a frequency range of 100kHz -- 2.5GHz, with a gain of 25dB at 500MHz
and noise figure of 2.4dB at 500MHz. Finally, the output of the amplifier
is passed through a 48MHz hardware low-pass filter. Our setup is situated in a
normal office environment, not in an EM-clean room.
\footnotetext{No data sheet for this particular chip is available.
However, data sheets for the TC11IB do exist, and the period of manufacture
for this chip indicates it may be related to the TC11IA.
}
Capturing is done
with a PicoScope 3207B set to a 100mV range and 1GS/s capture rate at
an 8-bit resolution.

\subsection{Locating The User Program}
\label{sec:locateprogram}
Our PLC OS is not equipped to emit a trigger as described in
section~\ref{sec:triggers}.
When faced with this issue, we first verified that the alternative of
waveform matching works. However, we also concluded that our
analysis for layer 2 would be easier if we could indeed trigger the oscilloscope
instead of searching the entire waveform.

One solution we have tried to achieve this is waveform triggering.
This uses the waveform matching approach, but with a dedicated, relatively
inexpensive low-speed oscilloscope that constantly scans the waveform for a
pattern and generates a trigger signal when it finds a match.
Two devices that implement this are Riscure's icWaves~\cite{icwaves} and KU
Leuven's waveform matching trigger~\cite{Beckers2016}.
We had access to an icWaves, and we
managed to produce a reliable and stable trigger signal based on the transition
from the operating system to the user program.
One issue we encountered was that having two
oscilloscopes on the same signal line with a T-splitter causes artefacts in
the measurements, causing us to abandon this approach.
We have not explored this further, but it could be remedied
by using a second probe.

We decided next on trying to emulate an OS trigger, by sending it from the
user program. As mentioned in section~\ref{sec:PLCs}, the software on the PLC
performs a read-execute-write-cycle. Since we are not interested in analysing
either the read or the write cycle, we can consider only part of the
user program as
interesting, and treat other parts around it as though they were part of the OS.
We introduce empty operations around the interesting part, and
after these empty operations we add a toggle that toggles one
of the output LEDs of an I/O simulator module. We have soldered a pin
to the back of this LED, and hook up a normal current measuring probe to
the EXT port of the oscilloscope. We now have a rising / falling edge
trigger for the oscilloscope, and a clear demarcation of the part of the
user program intended for analysis. For real-world operation, such an invasive measure
is clearly not an option, but it does not detract from our analysis.
The resulting trigger is not perfect, and requires us to preprocess
the measurements before analysis, as described in
section~\ref{sec:construction}.


\begin{figure}[t]
	\centering
	\includegraphics[width=.9\textwidth]{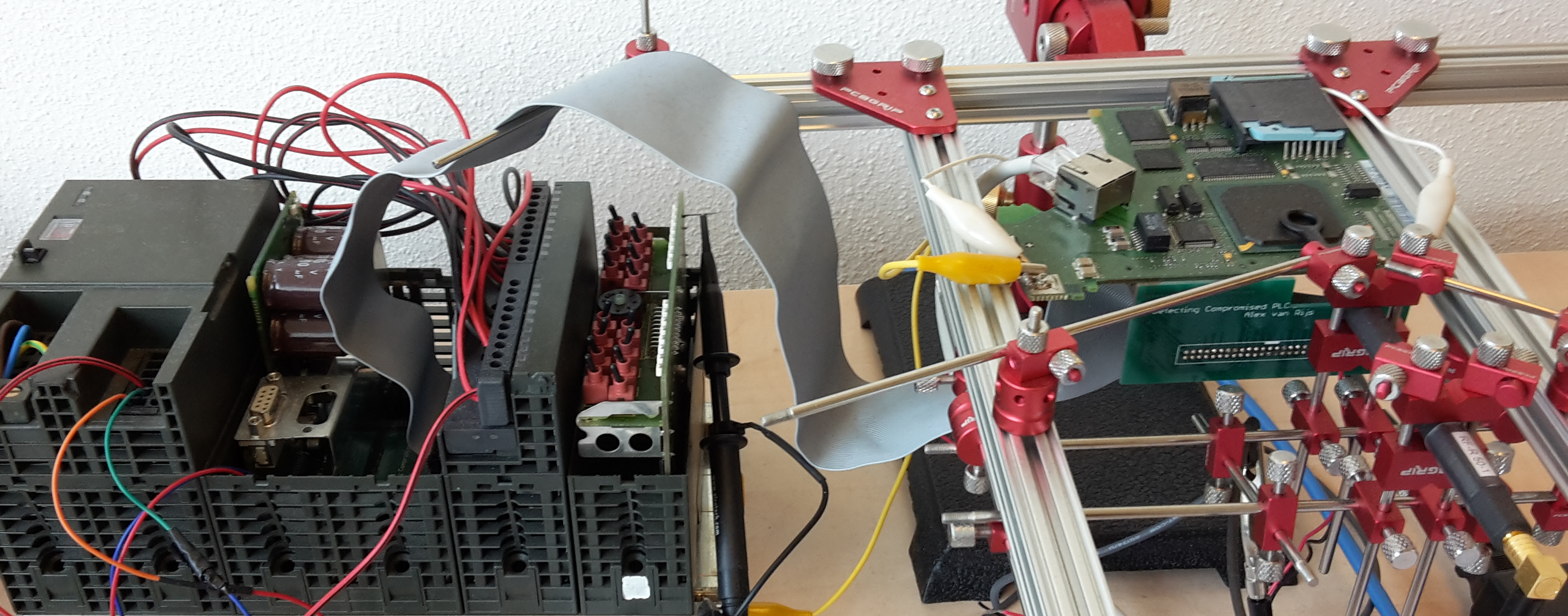}
	\caption{PLC with extruded mainboard and probe in place}
	\label{fig:plc}
\end{figure}

\subsection{Code Under Test}
\label{sec:code}
The Siemens S7-317 can be programmed in four different languages. Our analysis
focuses on one of these, SCL.

We initially attempted to make our analysis easier by eliminating branching
in the user program entirely, so that it would a single path through the program
that would take a constant amount of time and only deviate if different
instructions were executed.
However, this proved to be impossible:
first, because experimental results show that there are timing variations even
when the same instructions are executed on the same inputs
; and second, because
even simple programs like the one in listing~\ref{src:PrA} have multiple
paths through the program depending on their inputs.

The legitimate user program we want to recognize is given in listing~\ref{src:PrA}.
We will refer to this as program A, or PrA.
It is a very simple representation of a control system used to keep
a water level between two acceptable values, e.g. in a canal. Based on
whether the water level, simulated as a 4-bit input, is too low, too high,
or in-between, three different simulated outputs are driven. These outputs
could also be outputs to water pumps, warning lights, etc. Lines 1 and 2
read the water level input byte, compare it to the acceptable
levels, and set internal variables to indicate high or low water. Next,
line 3 uses these internal variables to determine whether this is or is not an
acceptable water level. This could obviously be done with a different
construction;
the current logic of inverting the XOR of the existing variables is a
result of the aforementioned attempt to achieve constant-time operation. We
have kept it since it lowers the number of comparisons and introduces additional
operations (NOT and XOR). Finally, on lines 4--6, the three
outputs are driven.


Next, we define two programs, PrB and PrC, that we want to
distinguish from PrA. These simulate slight changes that an adversary might
make to the program to influence its execution without influencing its runtime,
thereby evading layer 1 of our IDS. The changes are shown
in listings~\ref{src:PrB} and~\ref{src:PrC}. We have tested our method with
other programs with only minor changes, and the performance is similar. The
changes are:

\begin{itemize}
	\item In PrB, the attacker flips the logic of the \texttt{water\_low} variable,
		so that the system indicates low water when in fact, it is
		okay, or even high, and indicates okay when the water level is
		low. This simulates the attack where an attacker changes an
		instruction in the program code.

	\item In PrC, the attacker changes the constant in the comparison for
		\texttt{water\_high} to 12, so that the system potentially overflows
		without ever indicating anything other than an okay water level.
		This simulates the attack where an attacker changes only a
		comparison constant in the program code.

\end{itemize}


\begin{figure}[t]
\begin{center}
	\begin{minipage}{.48\textwidth}
\begin{lstlisting}[
		  basicstyle=\scriptsize\ttfamily,
		  basewidth=0.5em,
		  breaklines=true,
		  numbers=left,
		  stepnumber=1,
		  firstnumber=1,
		  caption=Program A,
		  label={src:PrA},
		]
#water_low := "DIGITAL_IN_CHAR" < CHAR#5;
#water_high := "DIGITAL_IN_CHAR" > CHAR#10;
#water_good := NOT (#water_low XOR #water_high);
"WATER_ADD_PUMP" := #water_low;
"WATER_OK" := #water_good;
"WATER_REMOVE_PUMP" := #water_high;
\end{lstlisting}
	\end{minipage}
	\begin{minipage}{.48\textwidth}
		\begin{minipage}{\textwidth}
\begin{lstlisting}[
			  basicstyle=\scriptsize\ttfamily,
			  basewidth=0.5em,
			  breaklines=true,
			  numbers=left,
			  stepnumber=1,
			  firstnumber=1,
			  caption=Changes in Program B,
			  label={src:PrB},
			]
#water_low := "DIGITAL_IN_CHAR" > CHAR#5;
\end{lstlisting}
		\end{minipage}
		\begin{minipage}{\textwidth}
\begin{lstlisting}[
			  basicstyle=\scriptsize\ttfamily,
			  basewidth=0.5em,
			  breaklines=true,
			  numbers=left,
			  stepnumber=1,
			  firstnumber=2,
			  caption=Changes in Program C,
			  label={src:PrC},
			]
#water_high := "DIGITAL_IN_CHAR" > CHAR#12;
\end{lstlisting}
		\end{minipage}
	\end{minipage}
\end{center}
\end{figure}

\section{Intrusion Detection Results}
We have described how an adversary can alter the code with
minimal impact on the program timing in section~\ref{sec:code}. Since this
evades layer 1 of our IDS, in the next sections we will
discusses the techniques applied for layer 2. We explain the steps we
took to prepare the captured dataset for analysis, the different analysis
techniques used, and the accuracy we achieved with these techniques.

\subsection{Template Construction}
\label{sec:construction}
The dataset captured from the Siemens S7-317 contains small interrupts,
variability in instruction execution time and clock jitter. These all cause
trace misalignment. To correct for this, we align the traces at the beginning
of the user program and filter out those traces where the user program has been
severely altered by interrupts. This filters out roughly 10\% of traces.
As mentioned in section~\ref{sec:locateprogram}, for the purpose of our analysis
we can treat the start and end of the user program as though they are part of
the OS. We ensure that these parts are areas of low EM emissions, and use a
peak finding algorithm to align on the first high peak after a valley: the part
of the user program being analysed.

We build profiles, or templates,
for user programs in several different ways, using progressively more
complex and more informative statistical models. Our aim is to test the accuracy
of such models in the intrusion detection context and identify the best model
for layer 2. For every chosen model we answer questions 1--4 posed in
section~\ref{sec:twolayer}, and show their performance for question 4, the
recognition problem, via the Receiver Operator Characteristic
(ROC), False Accept Rate / False Reject Rate (FAR/FRR), and Kernel Density
Estimation curves\footnotemark.

\footnotetext{
The ROC curve shows how, as the rate of genuine accepts (GAR) increases, the
rate of false accepts (FAR) increases as well. An ideal system has a 100\% GAR
with a 0\% FAR, and a perfect ROC curve looks like the one in
figure~\ref{fig:resultplot-combined-average-template-normal-logic-switched}.
The FAR/FRR curve shows the balance between the two error counts, and the
intersection in the graph denotes the Equal Error Rate (EER): it indicates the
threshold where the FAR is equal to the FRR, and is a good indication of the
accuracy of the system.  An EER of 50\% is bad performance, an EER of 0\% is
perfect.
For illustration purposes, we also include the kernel density estimation plots of
the scores for the genuine user program and the manipulated user program. The
more overlap these kernels have, the harder it is to recognize one as genuine
and the other as compromised.
}

We commence our analysis creating templates based on average and median traces,
i.e. we partition our experimental data in training and test sets and compute
the mean and median trace vectors using the training set. Template matching with
the test sets is performed using Sum of Absolute Differences (SAD) and
cross-correlation (XCORR) as distinguishing metrics.

Continuing, we also construct full side-channel
templates~\cite{DBLP:conf/ches/ChariRR02}. We assume that the EM leakage
$\mathbf{L}$ can be described by a multivariate normal distribution, i.e.
$\mathbf{L} \sim \mathcal{N}(\mathbf{m},\mathbf{\Sigma})$ with mean vector
$\mathbf{m}$ and covariance matrix $\mathbf{\Sigma}$, that are estimated using
the training set. Specifically, for every program PrI, I $\in$ \{A,B,C\}, we
estimate the parameters of the distribution $(\mathbf{L} |$ PrI $)$, and template
matching  is performed using a maximum likelihood approach. The EM leakage
$\mathbf{L}$ contains a large number of samples (in the range of several
thousands), requiring a high data complexity for the sufficient training of the
multivariate model. Thus, we rely on dimensionality reduction techniques such as
linear discriminant analysis (LDA)~\cite{DBLP:conf/ches/ArchambeauPSQ06} in
order to compress the traceset and select the most informative samples, often
referred to as Points of Interest (POIs).


\begin{figure}[t]
	\centering
	\vspace{-.5em}
		\includegraphics[width=.9\textwidth]{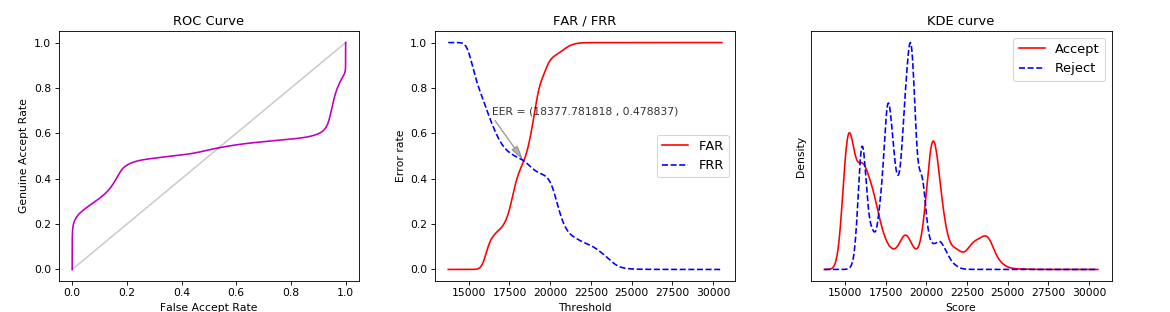}
	\caption{SAD results for a combined average trace of PrA compared to PrB}
	\label{fig:resultplot-combined-average-normal-logic-switched}
\end{figure}

\begin{figure}[t]
	\centering
	\vspace{-.5em}
		\includegraphics[width=.9\textwidth]{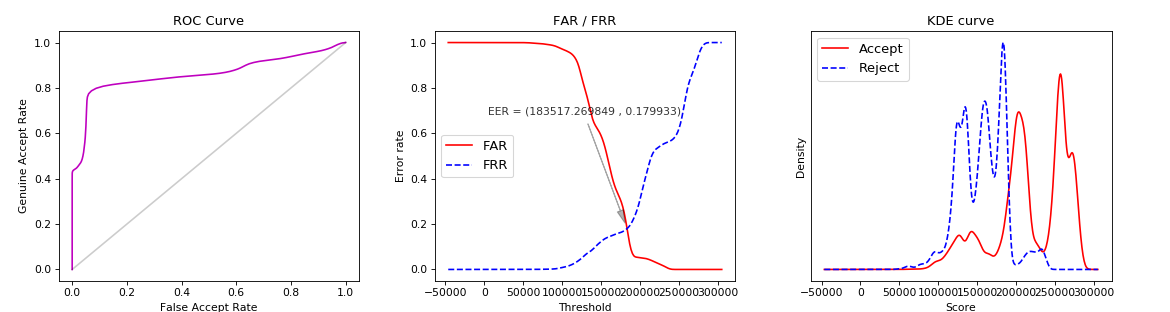}
	\caption{XCORR results for a combined average trace of PrA compared to PrB}
	\label{fig:resultplot-combined-average-xcorr-normal-logic-switched}
\end{figure}

\subsection{Averages and Medians with SAD and XCORR}
%
To answer question 1, ``can we distinguish PrA from PrB, and PrA from
PrC?'', we have built an average of all paths taken for every input of
the entire program for PrA, PrB, and PrC. For distinguishing PrA and
PrB, this works unexpectedly well; both Sum of Absolute Differences (SAD) and
cross-correlation (XCORR) manage to reach an 85\% recognition rate, i.e. for
both programs, 85\% of their traces are correctly identified as belonging to that
program.
For distinguishing PrA and PrC, however, PrA is only matched for 60\% of its
traces, and PrC is only matched for 50\% of its traces, with XCORR performing
slightly worse than SAD.

For question 2, ``can we distinguish between different paths in the \emph{same}
program?'', we have built averages of every input for PrA. When only accepting a
match if the exact input for each trace is matched, both SAD and XCORR perform
very badly, with a match rate lower than 20\%.
Since multiple inputs lead to the same path, we change our analysis to accept a
match if any of the inputs for that path match a certain trace. This improves
the accuracy significantly, with SAD reaching 93\%, and XCORR reaching 87\%.

For question 3, ``can we distinguish paths in user PrA from paths in PrB'', this
shows a combined behaviour from questions 1 and 2: distinguishing rates increase
as we accept paths, rather than specific inputs; and distinguishing between PrA
and PrB performs better than between PrA and PrC.

\begin{figure}[t]
	\centering
	\vspace{-.5em}
		\includegraphics[width=.9\textwidth]{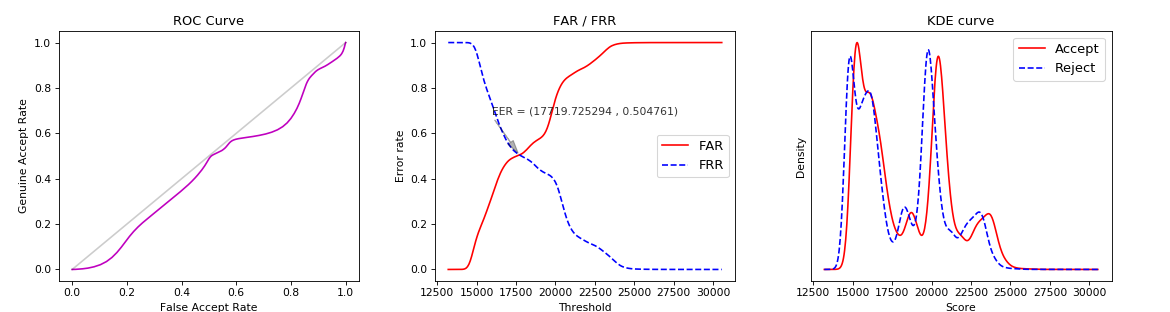}
	\caption{SAD results for a combined average trace of PrA compared to PrC}
	\label{fig:resultplot-combined-average-normal-high-flipped-12}
\end{figure}

For question 4, ``recognizing PrA'', SAD with an averaged trace for all inputs
on PrA performs very badly.
Figure~\ref{fig:resultplot-combined-average-normal-logic-switched}
shows the performance of this method when using it for the changed instruction
in PrB. Important to note is the overlap between
the estimated kernels in the results. The dotted graph is the set that should be
rejected, the unbroken one is the set that should be accepted. The overlap in
SAD scores shows that this algorithm simply is not good enough to distinguish between
variation from changing instructions and variation inherent in a single program with
multiple execution paths.
Using a combined median trace does not significantly change the performance of
the SAD method.
However, XCORR does perform rather well for recognizing PrB as not being PrA on
a combined average traceset, as can be seen in
figure~\ref{fig:resultplot-combined-average-xcorr-normal-logic-switched}.
The equal error rate is 18\%, significantly better than the 48\% that SAD
achieves here.
For the case of recognizing PrC as not being PrA, however, both XCORR and SAD
perform badly, achieving an EER of 50\%.
Figure~\ref{fig:resultplot-combined-average-normal-high-flipped-12} shows the
graphs for SAD.

Thus, we conclude that SAD is useless for the recognition problem, and although
XCORR can be used to recognize instruction changes, it cannot be used to recognize
comparison constant changes.

\subsection{Multivariate Templates}
The results of multivariate templating show significant improvements upon
the simpler models. For question 1, using LDA and only 10 POIs we get a perfect
distinguishing rate between both PrA and PrB, and PrA and PrC, when combining
all the inputs in a single dataset to train on.

However, for question 2, when
taking every input as a separate template, the performance degrades
significantly. Using an increased amount of POIs, attack traces and the improved
performance formulas of Choudary et al.~\cite{DBLP:conf/cardis/ChoudaryK13}, the
correct distinguishing rate for many inputs does not exceed 25\%, indicating the
need for a more detailed training phase. If we
combine the different inputs for the same path into a single template, however,
the distinguishing rate improves again.

For question 3, we see that distinguishing between paths for the same program
functions well if only a single path for each program is considered. When
multiple paths for each program are templated, the same effect we saw in
question 2 degrades the results.

\begin{figure}[t]
	\centering
	\vspace{-.5em}
		\includegraphics[width=.9\textwidth]{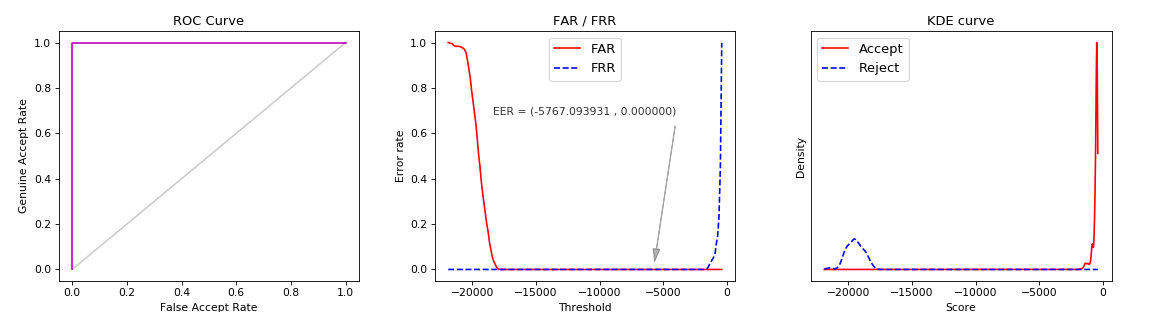}
	\caption{Template results for PrA compared to PrB}
	\label{fig:resultplot-combined-average-template-normal-logic-switched}
\end{figure}

\begin{figure}[t]
	\centering
	\vspace{-.5em}
		\includegraphics[width=.9\textwidth]{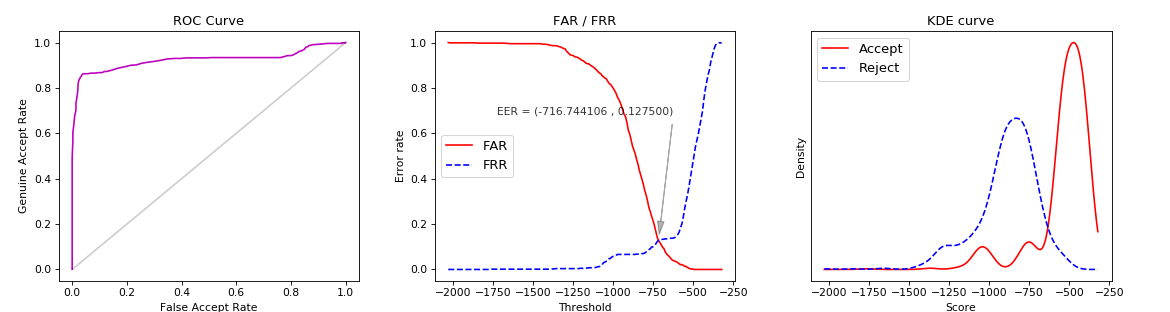}
	\caption{Template results for PrA compared to PrC}
	\label{fig:resultplot-combined-average-template-normal-high-flipped}
\end{figure}

However, for an IDS, question 4 is the most important one, and multivariate
templates do perform very well for this.
The best method we have found is to combine all the
traces for a single program into a single template, which relates to question 1.
For recognizing PrA with the attack of PrB, the changed instruction,
we get a perfect acceptance and rejection rate, with a very broad margin to set
the threshold. This can be seen in
figure~\ref{fig:resultplot-combined-average-template-normal-logic-switched}.
The broad margin indicates that changing an instruction is easily detected
by multivariate templating.
However, recognizing PrA
with the attack of PrC shows that the scores when changing only a comparison
constant are very close together.
Still, where both SAD and XCORR were unable to recognize PrA in the presence of
PrC, full templates are able to perform with a 13\% equal error rate, as shown
in figure~\ref{fig:resultplot-combined-average-template-normal-high-flipped}.

\section{Discussion \& Future Work}
%
Our results indicate that our IDS is capable of detecting very minor program
alterations through the use of full templating of EM emissions. We stress that
simple models such as sum of absolute differences and cross-correlation are
incapable of detecting the same alterations, so multivariate techniques are a
de facto requirement against detection-aware attackers. However, even with
multivariate templating, we note that the recognition threshold has a much
narrower margin for the most subtle attack of changing a comparison constant;
as can be seen by comparing the distance between kernels in the KDE plots of
figures~\ref{fig:resultplot-combined-average-template-normal-logic-switched}
and~\ref{fig:resultplot-combined-average-template-normal-high-flipped}.
Future work could expand to other classification techniques, including unsupervised
machine learning, to improve these recognition rates.

Our proposed IDS focuses on the user
program, because it is rather stable and can be treated as a grey box. We
do have access to the source code, if not to the specific hardware designs and
machine code. The operating system, however, remains a black box to us,
introducing interrupts, unpredictability of network communications, etc. Thus,
future work could look into profiling the normal behaviour of these PLCs,
including operating system operation, interrupts, and timing variations.
Similarly, more complex user programs
could be considered. As the numbers of possible inputs and control flows increase,
potential program behaviours become prohibitively numerous. For more complex user
programs, then, our technique could be applied to smaller units, like functions,
with another method to verify that these are executing in an expected order.

Our analysis is performed on programs written in SCL. However, as mentioned in section~\ref{sec:code},
the Siemens S7-317 can also be programmed in three other languages.
These three languages provide the same functionality to the
programmer, and all three are converted into STL before being uploaded to the device.
STL, short for Statement List, is Siemens' implementation of the IEC 61131-3 language
Instruction List, a low-level language resembling assembly. However,
when executing user programs based on STL on the PLC, a just-in-time (JIT)
compilation seems to occur. The first execution
of an STL-block in a user program run produces a longer and different
waveform from subsequent executions in the same user program run.
Future work can look into dealing with this JIT compilation and STL.

We stress that the actual
deployment of our side-channel IDS is not trivial. The main hindrance is
template transferability~\cite{DBLP:conf/eurocrypt/RenauldSVKF11}, i.e. the
fact that we can only train our statistical models on a limited amount of
devices, yet the model needs to be representative of a larger population of
devices. Even devices of the exact same model exhibit electrical variations due
to ageing and different manufacturing techniques, thus limiting the
effectiveness of our detection process.
On top of that, PLCs are often deployed in environments rich in EM-noise, which
may negatively impact our analysis. We did not have access to such an
environment, but it should be noted that our setup was in an office building,
not an EM-clean room. Also, the particular sensor we used seemed more sensitive
to noise coming from the chip than from the environment.

Another important consideration for deployment is whether the system being tested
can be disconnected from
its controlled process for the duration of the test. Since the user program behaviour
should depend on its inputs, this way the operator can
verify all expected paths are still present. The concern here
is that a potential attacker may simply remove the fail-safe code path, but leave
the conditional check on whether it should be taken in place. Since no additional
code is executed, nor any code normally executed is removed, the behaviour of the
program stays the same. Unfortunately, for most applications of PLCs, it is not
feasible to stop the industrial process being controlled or disconnect the PLC,
to check for this attack.

The final hindrance we wish to highlight here is cost: fitting a large amount
of legacy systems with EM probes would require a significant investment of
engineering time and money.

These issues combined may make it infeasible to deploy our system for anything
but the most critical systems. Future work can aim towards effective deployment
of high-accuracy side-channel IDSs, and analyse the effect of environmental
noise in detail.

\section{Conclusion}
In our work, we have shown that through time- and EM-monitoring techniques it
is possible to distinguish between user programs on programmable logic
controllers. This severely limits attackers and forces them to apply more
advanced techniques than naively replacing the user program.  In addition, we
have demonstrated that even a detection-aware adversary making very small
modifications to an existing user program can be effectively detected through
the use of full templating of EM emissions.
We have proposed an IDS for industrial control systems based on these techniques,
and demonstrated its feasibility for systems where only limited knowledge of the
platform and exact software instructions running on it is required.

To the best of our knowledge, we are the first to propose and demonstrate the
possibility of using the EM side-channel for this type of IDS on industrial
control systems.

All software created in the course of this research is made freely available to the
extent possible under applicable law at \url{https://polvanaubel.com/research/em-ics/code/}.

\begingroup
\setlength{\emergencystretch}{8em}
\printbibliography
\endgroup

\end{document}